\documentclass[%
 reprint,
 amsmath,amssymb,
 aps,
]{revtex4-1}

\usepackage{graphicx}
\usepackage{dcolumn}
\usepackage{bm}

\begin{document}

\preprint{APS/123-QED}

\title{Nuclear Dynamics at the Particle Threshold}

\author{Nadia Tsoneva}%
\email{tsoneva@theo.physik.uni-giessen.de}
 \altaffiliation[Also at ]{INRNE, 1784 Sofia, Bulgaria}
\author{Horst Lenske}
 \affiliation{Institut f\"ur Theoretische Physik, Universit\"at Gie\ss{}en}

\date{\today}

\begin{abstract}
Recently, new low-energy modes of excitation called pygmy resonances have been observed. Their distinct feature is the close connection to nuclear skin oscillations. A successful description of the pygmy resonances could be achieved in a microscopic theoretical approach which incorporates the density functional theory and  QRPA formalism extended with multi-phonon degrees of freedom.  The latter is found of crucial importance for the understanding of the fine structure of nuclear electric and magnetic excitations at low energies. Corresponding microscopic multi-phonon response functions are implemented in the studies of s-process of nucleosynthesis.
\end{abstract}

\maketitle


\section{Introduction}

Leaving the valley of stability, nuclear dynamics is changing gradually as indicated by the appearance of new modes of excitations. Such examples are the observation of halos in light nuclei \cite{Tan85} and the skin phenomena in medium and heavy nuclei \cite{Suz90,Isa92} closely connected with the isospin dynamics of nuclear matter. Recent development of experimental facilities on radioactive nuclear beams opens the opportunity to investigate unknown regions of exotic nuclei.  
One of the most interesting findings, was the observation of enhanced dipole strength close to the particle emission threshold as a common feature of stable and unstable nuclei with neutron excess \cite{Sav13,Aum05,Adr05,Vol06,Sch08}. It was associated with oscillations of a small outer layer of neutron-rich nuclear matter with respect to the isospin symmetric nuclear core. As this clustering of mostly neutron single-particle electric dipole transitions exhausts only a minor portion of Thomas-Reiche-Kuhn (TRK) sum rule it was named pygmy dipole resonance (PDR).
The mainly electric character of the PDR was explained in recent investigations of low-energy E1 and
spin-flip M1 excitations in N=82 and N=50 nuclei \cite{Ton10,Schw13} which allow to decompose the dipole strength below the GDR to elastic E1 component, related to PDR skin
oscillations and background components composed of elastic and inelastic
E1 and M1 transitions, respectively.

An obvious question, coming up immediately in this connection, is to what extent the presence of a neutron or proton skin will affect excitations of other multipolarities and {\em vice versa}. Promising candidates are low-energy 2$^{+}$ states, especially those in excess of the spectral distributions known from stable nuclei. Quadrupole response functions are investigated theoretically in neutron-rich Sn nuclei. A close connection of low-energy 2$^+$ excitations and nuclear skins is obtained. These quadrupole states are related to Pygmy Quadrupole Resonance (PQR) \cite{Tso11}.

Here, we present systematic theoretical studies, based on a method incorporating self-consistent Skyrme Hartree-Fock-Bogolubov (HFB) \cite{Hofmann} and Quasiparticle-Phonon Model (QPM) theory \cite{Sol76}, of dipole and other multipole excitations over isotonic and isotopic chains. We consider not only nuclei with neutron excess but also nuclei close to the proton drip line \cite{Vol06,Paa07,Sch08,Tso04,Tso08,Ton10,Tso11}.  
Our quasiparticle-random-phase approximation (QRPA) calculations indicate a correlation between the observed total PDR strength and the neutron-to-proton ratio $N/Z$  \cite{Tso08}.
From the analysis of transition densities, the unique behavior of the PDR  is revealed, making it distinct from giant dipole resonance (GDR).
In addition, it has been suggested that the PDR is 
independent of the type of nucleon excess (neutron or proton) \cite {Tso08}. For the tin isotopic chain a transition of the neutron PDR to a proton PDR is found when the N=Z region is approached.

\section{The Theoretical Approach}
The model Hamiltonian resembles in structure the traditional QPM model \cite{Sol76} but in detail differs in the physical content in important aspects as discussed in Ref. \cite{Tso04,Tso08}.
In this sense, the approach is able to describe the nuclear ground state properties like binding energies, neutron and proton root mean square radii and the difference between them defining the nuclear skin, and separation energies \cite{Tso08}. 
The model Hamiltonian is given by:
\begin{equation}
{H=H_{MF}+H_M^{ph}+H_{SM}^{ph}+H_M^{pp}} \quad .\label{hh}
\end{equation}
Here, $H_{MF}=H_{sp}+H_{pair}$ is the mean-field part. Different from the standard QPM scheme this part is obtained from self-consistent HFB theory \cite{Hofmann}. The $H_{MF}$ defines the single particle properties including potentials and pairing interactions for protons and neutrons, such that also dynamical effects beyond mean-field can be taken into account. That goal is achieved in practice by using fully microscopic HFB potentials and pairing fields as input but performing a second step variation with scaled auxiliary potentials and pairing fields readjusted in a self-consistent manner such that nuclear binding energies and other ground state properties of relevance are closely reproduced. 

$H_M^{ph}$, $H_{SM}^{ph}$ and $H_M^{pp}$ are residual interactions,
taken as a sum of isoscalar and isovector separable multipole and
spin-multipole interactions in the particle-hole and multipole
pairing interaction in the particle-particle channels. The model parameters are fixed either empirically \cite{Vdo} or by reference to QRPA calculation performed within the density matrix expansion (DME) of G-matrix interaction discussed in Ref. \cite{Hofmann}.

\begin{figure}[tb]
\centering
\includegraphics*[width=130mm]{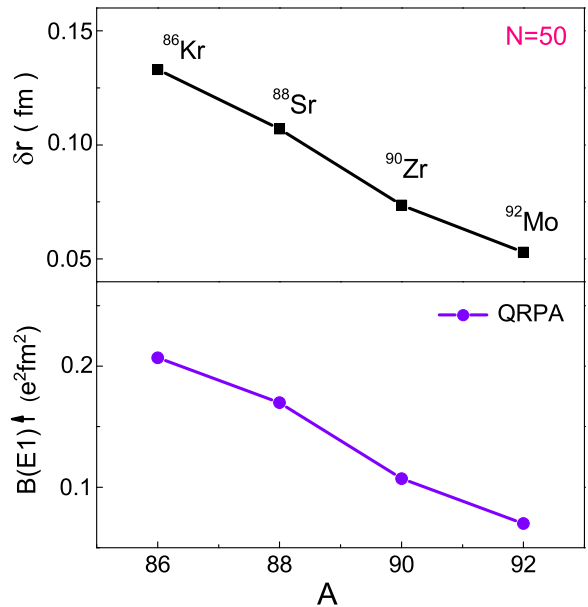}
\caption{(color online)
Total PDR strength obtained from QRPA calculations in comparison with the neutron skin thickness calculated with equation (5) in N=50 nuclei. Note, that the proton number Z increases with the mass number A.}
\label{Fig.1}
\end{figure}

\begin{figure*}[tb]
\centering
\includegraphics*[width=190mm]{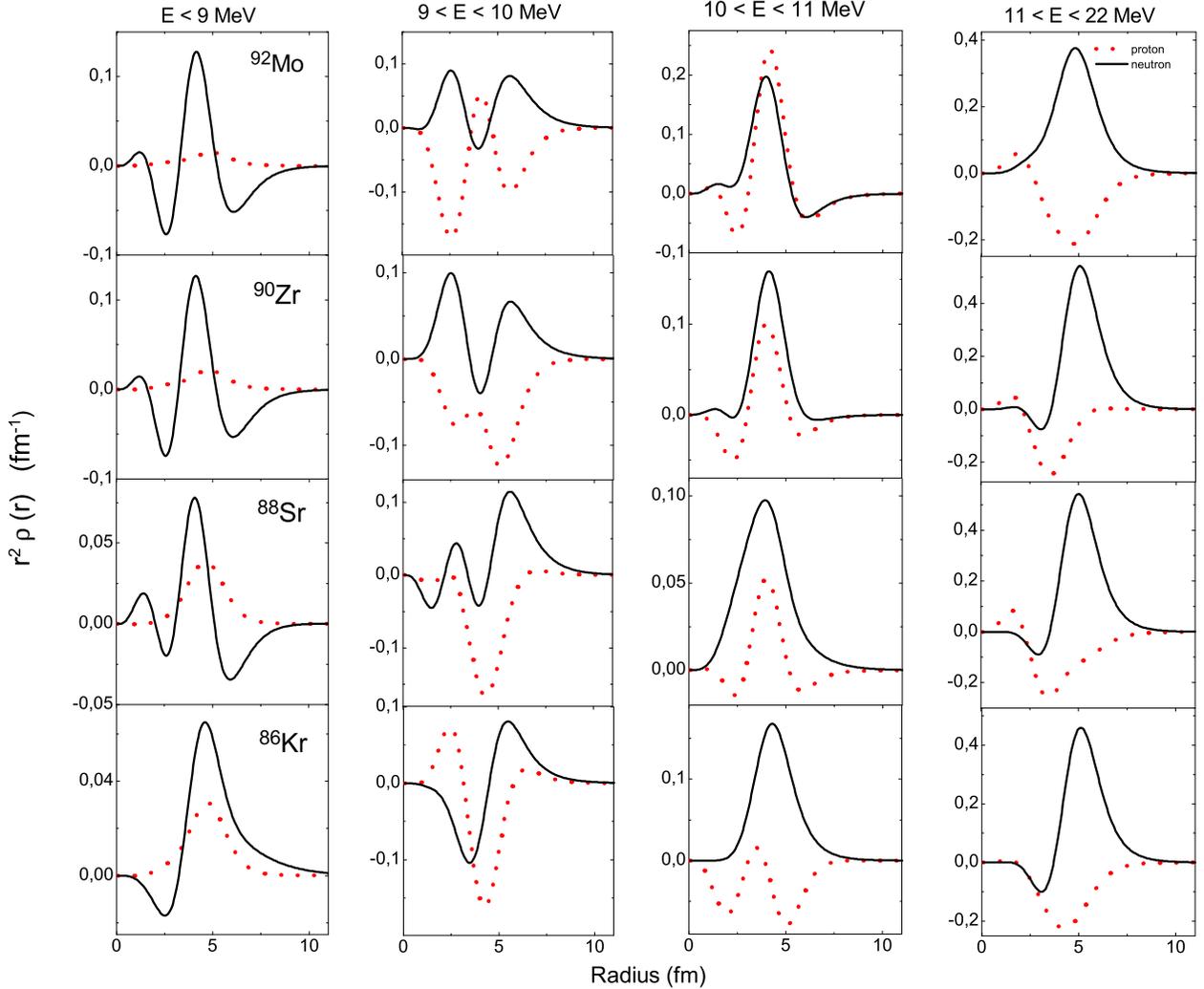}
\caption{(color online)
Dipole proton (in red) and neutron(in black) transition densities obtained from QRPA calculations in N=50 nuclei.}
\label{Fig.2}
\end{figure*}

 \subsection{The Nuclear Ground State}
The reliable description of ground state properties is of genuine importance for extrapolations of QRPA and QPM calculations into unknown mass regions. Taking advantage of the
Kohn-Sham theorem \cite{HoKohn:64,KohnSham:65} of Density Functional Theory (DFT) the total
binding energy $B(A)$ of the nucleus could be expressed as an integral over an energy
density functional with (quantal) kinetic ($\tau$) and self-energy
parts, respectively,
\begin{equation}
B(A)=\int{d^3r\left( \tau(\rho)+\frac{1}{2}\rho
U(\rho)\right)}+E_{pair}
\end{equation}
\[
=\sum_j{v_j^2\left(e_j-<\Sigma>_j+\frac{1}{2}<U>_j\right)}+E_{pair}
\quad ,
\]
and pairing contributions are indicated by $E_{pair}$. The second
relation is obtained from Hartree-Fock-Bogoliubov (HFB) theory with 
occupancies $v^2_j$ and
potential energies $<U>_j$ of the occupied levels $j$, see e.g.
\cite{Hofmann}. Above, $U(\rho)$ is the proper self-energy,
i.e. not including the rearrangement contributions from the
intrinsic density dependence of nuclear interactions
\cite{Hofmann}. Hence, $U(\rho)$ has to be distinguished
from the effective self-energy obtained by variation
\begin{equation}
\Sigma(\rho)=\frac{1}{2}\frac{\partial \rho U(\rho)}{\partial
\rho}
\end{equation}
and appearing in the single particle Schroedinger equation. In
order to keep the QPM calculations feasible we choose $\Sigma
\equiv U_{WS}$ to be of Wood-Saxon (WS) shape with adjustable
parameters. By inversion and observing that the densities and
potentials in a finite nucleus are naturally given parametrically
as functions of the radius $r$, we find
\begin{equation}
\rho(r) U(r)= -2 \int_r^\infty{ds \frac{\partial \rho(s)}{\partial
s} U_{WS}(s)} \quad .
\end{equation}
Evaluating these relations with the microscopic proton and neutron
densities obtained by solving the Schroedinger equation with
$U_{WS}$ the potential $U(\rho)$ is the self-consistently derived
reduced self-energy entering e.g. into the binding energy.
 
In practice, for a given nucleus of mass $A$ the depth of the
central and spin-orbit potentials, radius and diffusivity
parameters of $U_{WS}$ are adjusted separately for protons and
neutrons to the corresponding single particle separation energies,
the total binding energy \cite{Audi95}, the charge radii and
(relative) differences of proton and neutron root-mean-square
(RMS) radii,
\begin{equation}\label{dr}
\delta r=\sqrt{<r^2>_n}-\sqrt{<r^2>_p} \quad ,
\end{equation}
from our previous HFB calculations \cite{Hofmann}.
In ref.\cite{Tso08} theoretically obtained RMS radii are compared to those determined from charge exchange reactions for a number of Sn isotopes. 
The approach sketched above leads to very
satisfactory results on binding energies and proton-neutron
RMS-differences as shown in \cite{Tso04,Tso08}. A smooth dependence of the parameters on $A$ is found which supports the reliability of the method.

Calculations of ground state neutron and proton densities for Z=50 and N=50, 82 nuclei are shown in \cite{Tso08,Sch08}. Of special importance for these investigations are the nuclear surface regions, where the formation of a skin takes place. A common observation
found in various investigated isotopic and isotonic chains of nuclei is that the thickness of the neutron skin correlates with the neutron-to-proton ratio N/Z.  For example, in Sn isotopes with A$\geq$106 the neutron distributions begin to extend beyond the proton density and the effect continues to
increase with the neutron excess, up to $^{132}Sn$. Thus, these
nuclei have a neutron skin. The situation reverses in $^{100-102}Sn$,
where a tiny proton skin appears. 

\subsection{The Nuclear Excited States} 

The nuclear excitations are expressed in terms of QRPA phonons:
\begin{equation}
Q^{+}_{\lambda \mu i}=\frac{1}{2}{
\sum_{jj'}{ \left(\psi_{jj'}^{\lambda i}A^+_{\lambda\mu}(jj')
-\varphi_{jj'}^{\lambda i}\widetilde{A}_{\lambda\mu}(jj')
\right)}},
\label{phonon}
\end{equation}
where $j\equiv{(nljm\tau)}$ is a single-particle proton or neutron state;
${A}^+_{\lambda \mu}$ and $\widetilde{A}_{\lambda \mu}$ are
time-forward and time-backward operators, coupling 
two-quasiparticle creation or annihilation operators to a total
angular momentum $\lambda$ with projection $\mu$ by means of the
Clebsch-Gordan coefficients $C^{\lambda\mu}_{jmj'm'}=\left\langle
jmj'm'|\lambda\mu\right\rangle$.
The excitation energies of the phonons and the time-forward and time-backward amplitudes
$\psi_{j_1j_2}^{\lambda i}$ and $\varphi_{j_1j_2}^{\lambda i}$ in Eq.~(\ref{phonon}) are determined by solving QRPA equations \cite{Sol76}.

Furthermore, the QPM provides a microscopic approach to multi-configuration mixing \cite{Sol76}. The wave function of an excited state consists of one-, two- and three-phonon configurations \cite{Gri94}: 
\begin{widetext}
\begin{equation}
\Psi_{\nu} (JM) =
 \left\{ \sum_i R_i(J\nu) Q^{+}_{JMi}
\right.
+ \sum_{\scriptstyle \lambda_1 i_1 \atop \scriptstyle \lambda_2 i_2}
P_{\lambda_2 i_2}^{\lambda_1 i_1}(J \nu)
\left[ Q^{+}_{\lambda_1 \mu_1 i_1} \times Q^{+}_{\lambda_2 \mu_2 i_2}
\right]_{JM}
\label{wf}
\end{equation}
\[
\left.
{+ \sum_{_{ \lambda_1 i_1 \lambda_2 i_2 \atop
 \lambda_3 i_3 I}}}
{T_{\lambda_3 i_3}^{\lambda_1 i_1 \lambda_2 i_2I}(J\nu )
\left[ \left[ Q^{+}_{\lambda_1 \mu_1 i_1} \otimes Q^{+}_{\lambda_2 \mu_2
i_2} \right]_{IK}
\otimes Q^{+}_{\lambda_3 \mu_3 i_3}\right]}_{JM}\right\}\Psi_0
\]
\end{widetext}
where R, P and T are unknown amplitudes, and $\nu$ labels the
number of the  excited states.

\begin{figure*}[tb]
\centering
\includegraphics*[width=180mm]{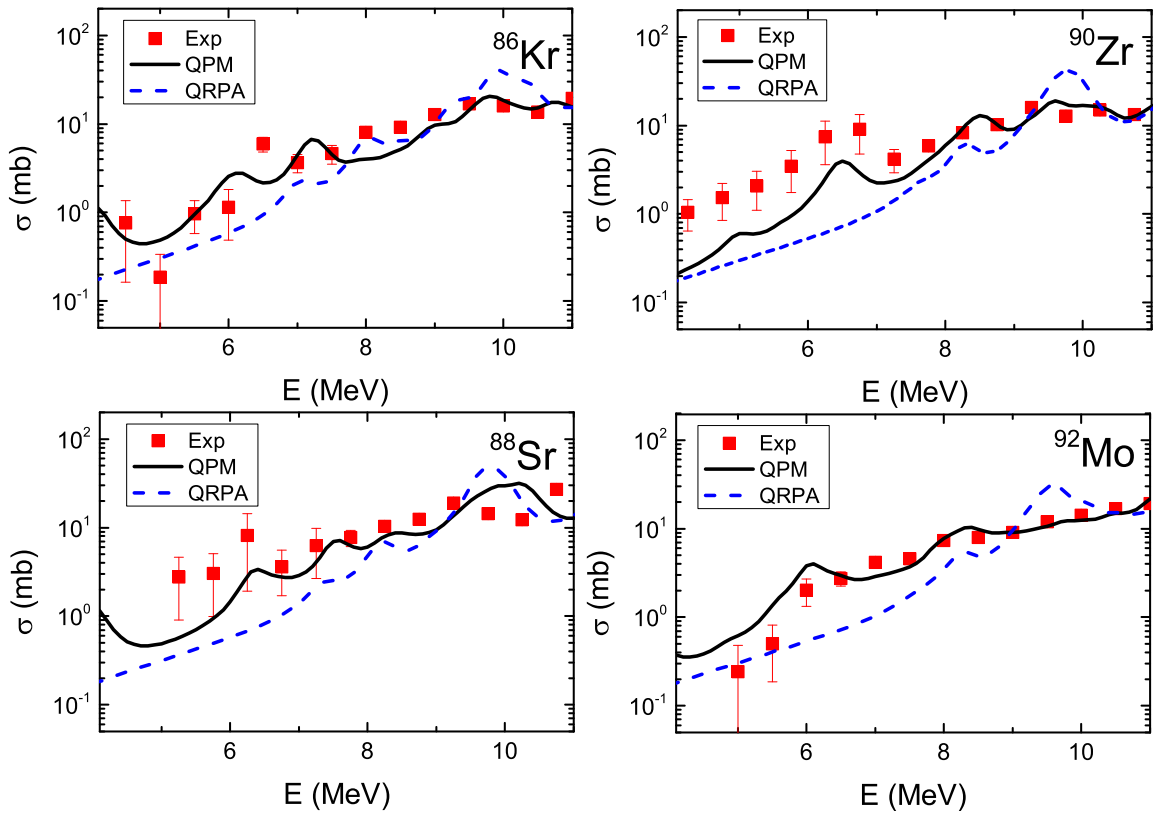} 
\caption{(color online)
N=50 nuclei: Systematic QRPA (in blue) and three-phonon QPM (in black) calculations of dipole photoabsorption cross section below the neutron threshold in comparison with data (in red) from [10].}
\label{Fig.3}
\end{figure*}

The electromagnetic transitions are described by transition operators accounting for the internal fermionic structure of the phonons \cite{Pon98}. The method allows for sufficiently large configuration spaces such that a unified
description of low-energy single and multiple phonon states and
the GDR is feasible. Such a unified treatment is exactly what is required in
order to separate the multi-phonon and the genuine PDR $1^{-}$
strengths in a meaningful way. 

\begin{figure}[tb]
\centering
\includegraphics*[width=102mm]{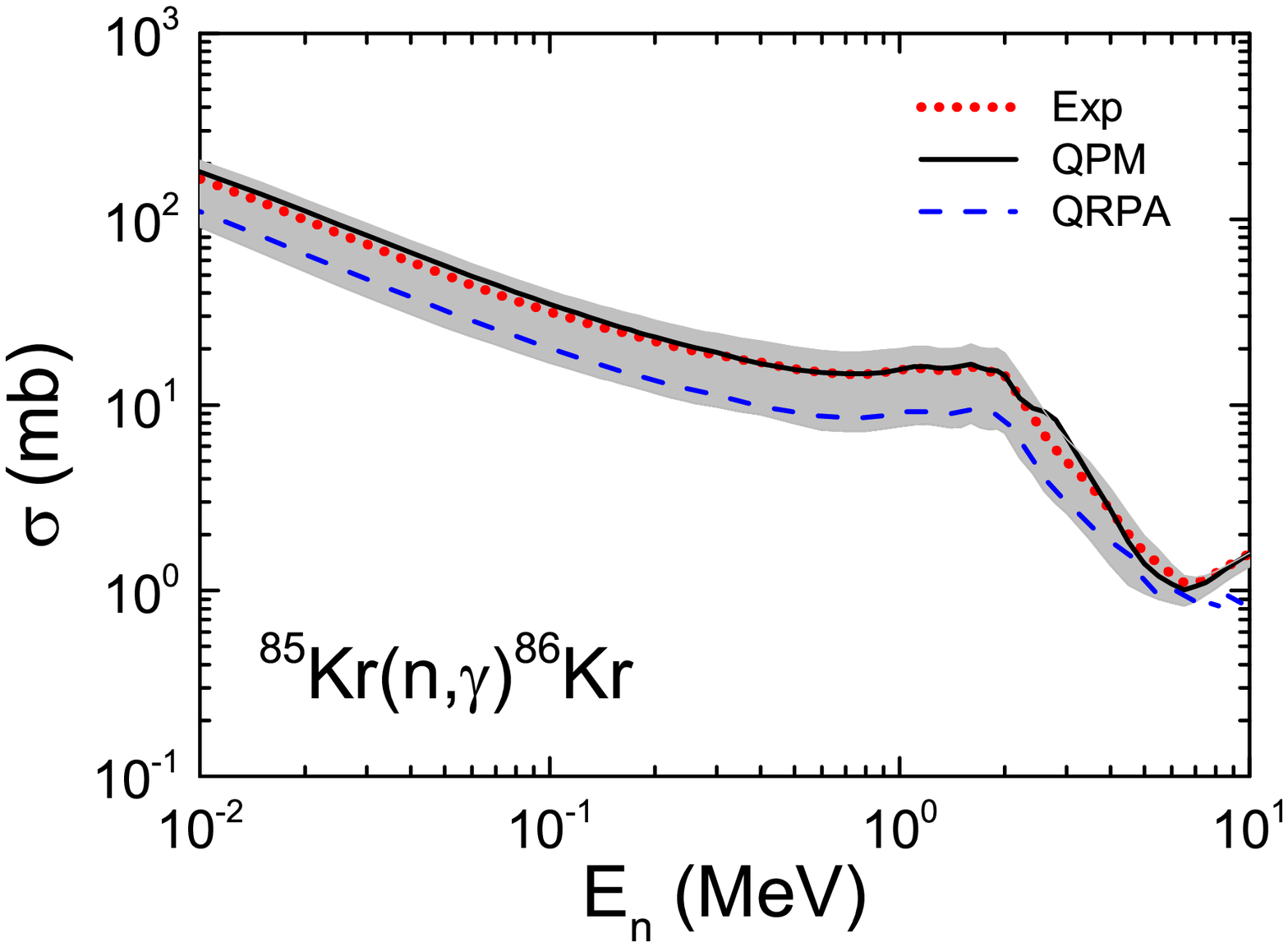} 
\caption{(color online)
Cross section of $^{85}Kr^g(n,\gamma)^{86}$Kr calculated with TALYS using experimental dipole (in red), QRPA (in blue) and three-phonon QPM strength functions (in black) from ref. [24]. The predicted uncertainties (shaded area) are derived from the experimental errors of the dipole strength function and from variations in the nuclear level density parameters.\\
}
\label{Fig.4}
\end{figure}

\section{Discussion}
\subsection{Spectroscopic studies of the dipole response from QRPA and three-phonon QPM calculations in N=50 isotones}
Systematic QRPA and QPM calculations of dipole strength functions in N=50 nuclei are compared to resent experimental results from \cite{Schw13}. 
Both, the experiment and the calculations reveal a strong enhancement of the  E1 strength in the excitation energy range E= 6$\div$10  MeV with respect to  the shape of a  Lorentz-like strength function used to adjust the GDR \cite{Schw13}.
From QRPA calculations shown in Fig. \ref{Fig.1} and in Fig. \ref{Fig.3}, the excitation energy region below E$<$9 MeV is related to PDR \cite{Tso08,Schw13} whose total strength smoothly decreases with increasing proton number Z. It is closely correlated with the thickness of the neutron skin which is presented in Fig. \ref{Fig.1}.

As the excitation energy is increased, the isovector contribution to the dipole strength increases following closely its Lorentzian fall-off often assumed with GDR in data analysis \cite{Ber75}. Theoretically, this can be seen in transition densities and state vectors structure which manifest an enlarging of the out-of-face neutron to proton contributions and corresponding energy-weighted sum rules which is generally associated with the GDR \cite{Tso08,Schw13}. QRPA dipole transition proton and neutron densities of N=50 nuclei are shown in Fig. \ref{Fig.2}. The excitation energy range E$<$9 MeV where the neutron oscillations dominate at the nuclear surface is related to a neutron PDR.
The analysis of the evolution of dipole transition densities with increasing excitation energy allows us to distinguish
between the PDR and other types of dipole excitations. In this aspect, the states in the excitation energy region of $9<$E$<10$ MeV carry a signature different from the PDR. The protons and neutrons start to move out of phase, being compatible with the low-energy part of the GDR. A strong argument in this direction is that the observed amount of dipole strength located in the excitation energy region $9<$E$<10$ MeV cannot be directly connected
to N/Z ratios and corresponding skins, as can be done for the dipole strength below 9 MeV. A further increase of excitation energy in the range of 10$<$E$<$11 MeV leads to dynamic processes of collective excitations of different neutron and proton subshells of the nuclear interior, which in some cases could be in-phase as shown in Fig. \ref{Fig.2}. It should be mentioned that such more or less collective excitations including a considerable contribution of inner-shell neutrons and protons should not be interpreted in the same way as the genuine PDR
mode explained by neutron skin oscillations. At excitation energy E= 11$-$22 MeV a strong isovector oscillation corresponding to the
excitation of the GDR is found.

However, we should point out that the QRPA is unable to account for the detailed description of the dipole strength function. This is clearly seen in Fig. \ref{Fig.3}. The comparison of QRPA and the three-phonon QPM calculations of dipole photoabsorption cross sections indicates that for the PDR region the coupling of QRPA PDR and GDR phonons and multiphonon states is very important. The result is a shift of E1 strength toward lower energies. 
This can be described in three-phonon QPM calculations which are able to reproduce the fine structure of the later fairly well as it follows from the comparison with the experiment presented in Fig. \ref{Fig.3}. Such precise knowledge of nuclear response functions is very important for the determination of photonuclear reactions cross sections for the astrophysics. 
\subsection{Determination of neutron-capture reaction cross sections related to the s-process of nucleosynthesis.}

The microscopic strength functions obtained from QRPA and three-phonon QPM calculations have been implemented into statistical reaction code to investigate neutron-capture cross sections of astrophysical importance. As an example case the neutron-capture cross section of the reaction $^{85}$Kr(n,$\gamma$)$^{36}$Kr \cite{Raut13} is studied. The $^{85}$Kr is a branching point nucleus of the s-process of the nucleosynthesis. The calculated $^{85}$Kr(n,$\gamma$)$^{36}$Kr cross section is shown in Fig. \ref{Fig.4} in comparison with experimental data from \cite{Raut13}.
It is seen that the neutron-capture cross section obtained from the three-phonon QPM is in a very good agreement with the experiment \cite{Raut13} while the QRPA gives a reduced value of about $\approx$ 35$\%$. The estimated value of the pure PDR contribution to the QRPA calculated neutron-capture cross section is of the order of $\approx$ 30$\%$. The agreement between the QPM theory and data is confirming the predictive power of the involved multiphonon theoretical methods for exploratory investigations of neutron-capture reaction rates in hitherto experimentally inaccessible mass regions. 

\subsection{Distinguishing of the Pygmy Dipole Resonance from other low-energy excitations}
A major experimental problem is to distinguish
$M$1 and $E$1 strength, since both of them are highly fragmented at these energies.
In order to unambiguously discriminate between these dipole excitation modes,
the spin and parity of the individual states must be known. That was achieved in high-sensitivity studies of E1 and M1 transitions observed in the $^{138}$Ba($\vec{\gamma}$,$\gamma$') reaction at energies below the neutron emission threshold which have been performed using the nearly monoenergetic and 100\% linearly polarized photon beams from the HI$\vec{\gamma}$S facility \cite{Pie01,Ton10}.
The electric dipole character of the so-called $pygmy$ mode was experimentally verified for excitations with energies E= 4.0 - 8.6 MeV \cite{Ton10}.
The fine structure of the $M$1 $spin-flip$ mode was observed for the first time in $N$ = 82 nuclei.
Three-phonon QPM calculations of low-energy E1 and M1 strengths in $^{138}$Ba \cite{Ton10} are in a good agreement with E1 and M1 data below the particle threshold, both with respect to the centroid energies and summed transition strengths. A common feature of the low-energy $1^-$ and $1^+$ states is that both modes are excited by almost pure two-quasiparticle (2-QP) QRPA states. They serve as doorway states which then decay into multi-configuration states with complicated multi-phonon wave functions, thus giving rise to fragmentation of the spectral distributions \cite{Ton10}.

Furthermore, in theoretical investigations of 2$^+$ excitations in Sn isotopic chain, we find a strength clustering of quadrupole states, at low-energies, of predominantly neutron structure. 
At the same time, the proton contribution to state vectors and B(E2) strengths, located in the  excitation energy range E$\approx$ 2-4 MeV, increases toward $^{104}$Sn and brings more intensive proton quadrupole excitations in $^{104}$Sn there. Consequently, the $^{104}$Sn nucleus appears to be an opposing case where a change from a neutron to a proton skin occurs.
Theoretical results of B(E2) strength distributions in $^{104,120,134}$Sn isotopes are presented in ref. \cite{Tso11}. A sizable increase of B(E2) strength at excitation energy E$\approx$ 2-4 MeV is observed for the heaviest tin isotopes - $^{130}$Sn and $^{134}$Sn.
The clustering of quadrupole states at low-energies shows a pattern similar to the PDR phenomenon. Therefore, we may consider the spectral distribution of a Pygmy Quadrupole Resonance (PQR). Correspondingly, the connection of the PQR with neutron or proton skin oscillations is demonstrated in the analysis of transition densities \cite{Tso11}. Similarly to the PDR, a transition from a neutron PQR to a proton PQR in $^{104}$Sn is observed for the mass region where the neutron skin reverses into a proton skin \cite{Tso08,Tso11}.

From the detailed analysis of the state vectors structure of the 2$^+$ excitations in Sn nuclei at excitation energies $E= 2-4 MeV$ we find that it is dominated by neutron 2QP excitations from the valence shells. The most important proton contribution in all isotopes is due to the ${[1g_{9/2}2d_{5/2}]}_{\pi}$ 2QP component, which, however, never exceeds 5$\%$ \cite{Tso11} which reflects the change of the proton binding energy $\epsilon_{b}$ of the $g_{9/2}$ level when approaching the $N=Z$ limit.
In the neutron sector, the contributions follow closely the evolution of the shell structure given in ref. \cite{Tso11}. In most cases the  $[2^{+}_{2}]_{QRPA}$ state vectors are dominated by re-scattering contributions related to re-orientation of the s.p. angular momenta. 
An effect of increased low-energy E2 strengths with increasing neutron number is observed as well in our QRPA calculations of lighter mass nuclei  $^{33}$Al/$^{35}$Al and $^{32}$Mg/$^{34}$Mg which are presented in ref. \cite{Tso14}. That mass region is of particular interest because of the well known 'island of inversion' at the N=20 shell closure \cite{Noc12}. 
Correspondingly, the connection of the low-energy 2$^+$ states with a PQR is demonstrated in the analysis of transition densities \cite{Tso11}. Strong neutron oscillations at the nuclear surface play a dominant role in the PQR energy range which is in agreement with Ref. \cite{Tso08}. Similarly to the PDR, a transition from a neutron PQR to a proton PQR in $^{104}$Sn is found for the mass region where the neutron skin reverses into a proton skin \cite{Tso11}. Furthermore, QPM calculations of $B(E2)$ and $B(M1)$ transition rates of low-lying 2$^+$ states in Sn nuclei indicate clearly differences from known collective states and scissors modes.

\begin{figure*}[tb]
\centering
\includegraphics*[width=84mm]{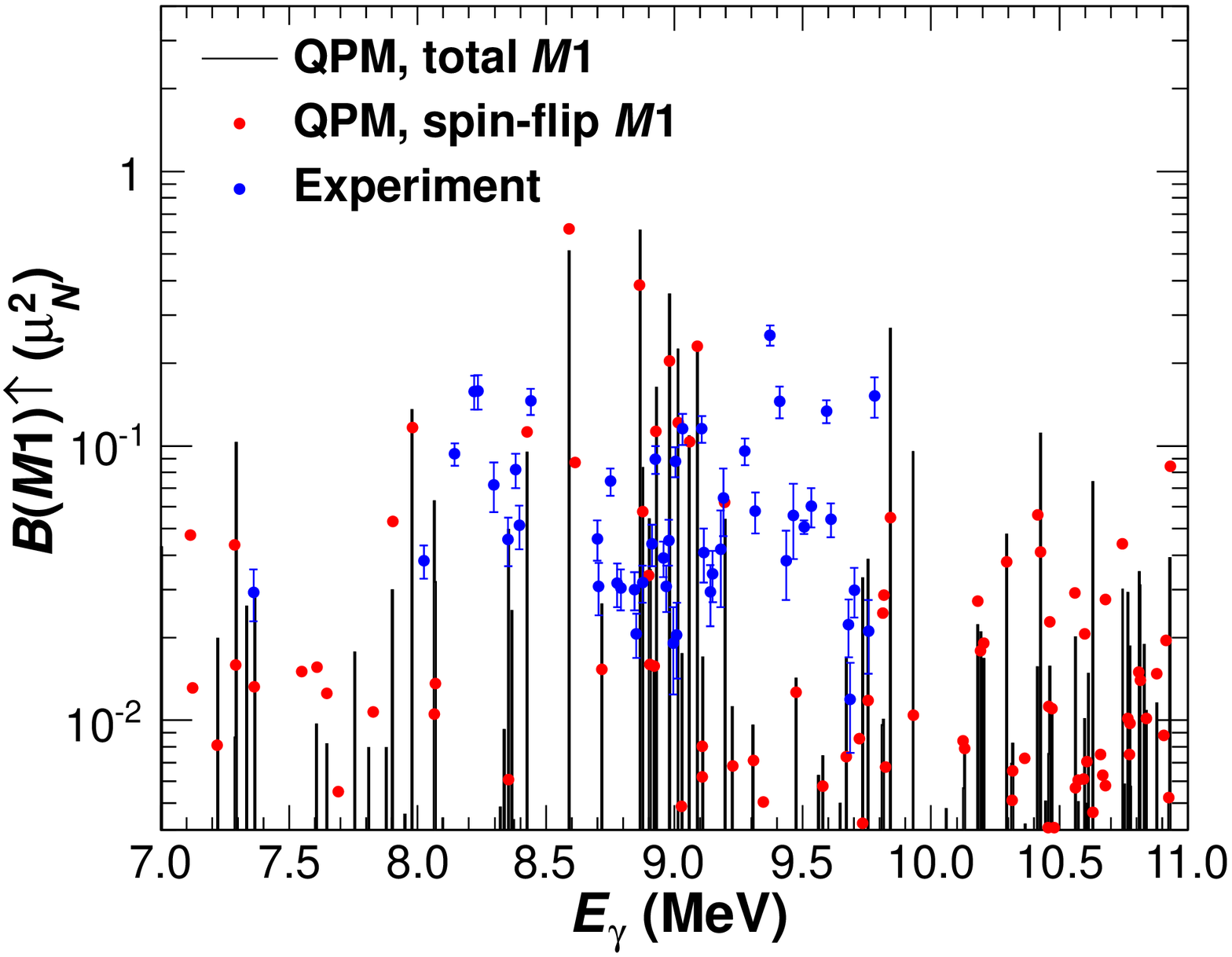} 
\includegraphics*[width=84mm]{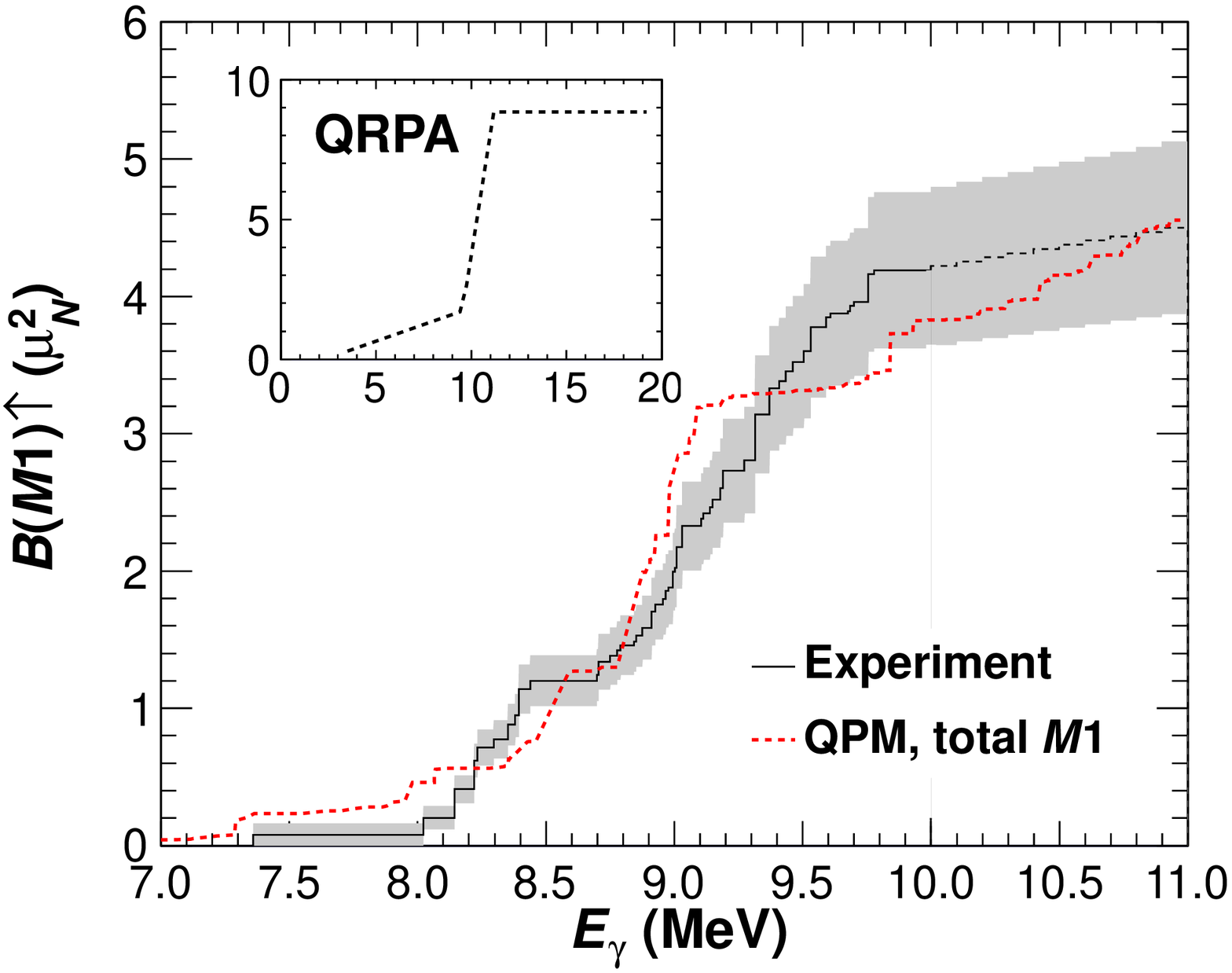} 
\caption{(color online) 
(left) The measured M1 distribution of discrete $1^+$ levels in $^{90}$Zr 
compared with three-phonon QPM predictions from ref. [28]. (right) Cumulative non-energy weighted M1 sum rule below E*=11 MeV.
}
\label{Fig.5}
\end{figure*}

\begin{figure}[tb]
\includegraphics*[width=99mm]{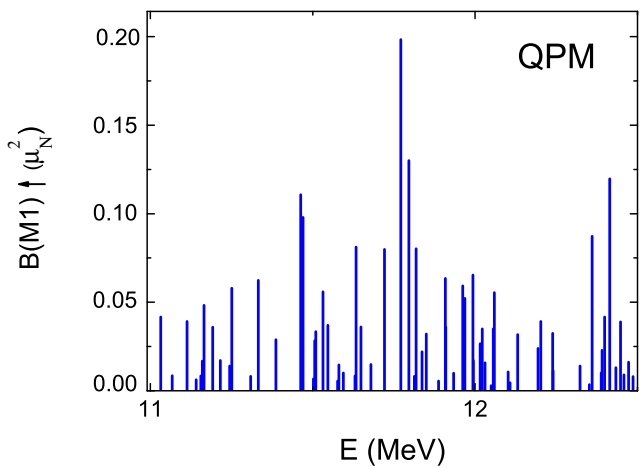} 
\caption{(color online) 
QPM prediction of M1 strength around the neutron threshold in $^{90}$Zr.
}
\label{Fig.6}
\end{figure}

\subsection{Description of the fine structure of the M1 spin-flip resonance in  $^{90}$Zr }

Recently, the fine structure of the $M1$-Giant Resonance (GR) in the nuclide $^{90}$Zr was investigated \cite{Rus13}. Measurements performed in the range 7-11 MeV reveal a $M1$ resonance structure with centroid energy of 9 MeV and a summed strength of 4.5(4) $\mu^2_N$. These data are fully reproduced in three-phonon QPM calculations \cite{Rus13}. The theoretical investigations which are presented in Fig. \ref{Fig.5} indicate a strong increase of the contribution of the orbital part of the magnetic moment due to coupling of multi-phonon states. 
Of special interest is the behavior of the M1 strength at higher energies close to and above the neutron-separation energy where the experimental accessibility is strongly reduced. For these regions, the theory predicts the existence of a strongly fragmented M1 strength with summed value of several $\mu_N^2$. 
Of special interest is the behavior of the
M1 strength at higher energies, namely in the range of 11 to 12.5 MeV. At these energies, which include the neutron-separation
energy (S$_n$=11.97 MeV), the experimental accessibility is strongly reduced. However, we can explore this region theoretically in the QPM by including one- and two-phonon configuration spaces with excitation energies up to E= 12.5 MeV. The model predicts a strongly fragmented M1
strength, related mainly to the decay of the 1$^+_4$ (QRPA) state into a considerable number of relatively uniformly distributed
1$^+$ states with very small transition probabilities, typically less than 0.2 $\mu_N^2$, and a total strength $\sum^{12.5MeV}_{11MeV}$ B(M1)$\uparrow\approx2\mu_N^2$. The later forms a tail-like extension of the giant M1 resonance around the neutron threshold which is shown in Fig. \ref{Fig.6}.
This is a very interesting finding which sheds light on the understanding of the long-standing problem with the quenching and dynamics of the M1 strength \cite{Rus13}. 

\section{Conclusions}
Investigations of low-energy excitations of different multipolarity
reveal new aspects on the isospin dynamics of the nucleus. In systematic investigations of low-energy excitations of different multipolarity in various isotopic and isotonic chains a specific signals of new modes of excitations related to PDR and PQR are observed. As a common feature, the structure of the PDR excited states in neutron-rich nuclei is dominated by neutron components while with the increase of the proton number, toward N=Z limit it transforms into a proton one. Consequently, the PDR transition strength is found directly related to the thickness of a neutron or proton skin driven by the Coulomb force. Its generic character is further
confirmed by related transition densities, showing that this mode
is clearly distinguishable from the conventional GDR mode.
In investigations of low-energy quadrupole states, PQR mode resembling the properties of the PDR is identified.
An interesting aspect is the close relation of the PQR excitations to the shell structure of the nucleus. The most convincing evidence for this feature is the disappearance of the PQR component in the double magic $^{132}$Sn nucleus. Thus, the PQR states are containing important information on the structure of the valence shells and their evolution with the nuclear mass number.

In these studies a common observation is that QRPA is unable to describe low-energy nuclear response functions in details. This can be achieved only if one takes into account the contribution of multi-phonon coupling. 
In three-phonon QPM calculations of N=82 nuclei the fine structure of the dipole strength below the neutron threshold is described as being mostly of electric character. Even though the magnetic  contribution to the PDR is small in comparison with the electric one it should be considered of a great importance for the understanding of the spin dynamics of skin nuclei. The theoretical investigations of the fragmentation pattern of the giant M1 resonance strength in $^{90}$Zr indicate a strong increase of the orbital contribution of the magnetic moment due to coupling of multiparticle-multihole configurations. The effect is estimated to account for about 22$\%$ of the total M1 strength below the threshold. Furthermore, a theoretical prediction of strongly fragmented M1 strength around the neutron threshold in $^{90}$Zr is done.

The agreement between data and calculations confirms the predictive power of the QPM many-body theory for exploratory investigations of new modes of excitation \cite{Tso08,Tso11,Tso04,Vol06,Schw13,Ton10,Rus13}. In this aspect, the  approach could be further developed and applied for investigations of hitherto experimentally inaccessible mass regions.

\begin{acknowledgments}
We wish to acknowledge the support of S. Goriely in providing us with
TALYS calculations and R. Schwengner for helping us with data analysis and useful discussions. The work is supported by BMBF grant 05P12RGFTE.
\end{acknowledgments}

\end{document}